\documentclass[12pt]{article}

\usepackage{amsfonts}
\usepackage{amssymb}

\def\hybrid{\topmargin -20pt    \oddsidemargin 0pt
             \headheight 0pt \headsep 0pt
             \textwidth 6.25in       
             \textheight 9.00in       
             \marginparwidth .875in
             \parskip 5pt plus 1pt   \jot = 1.5ex}

\hybrid

\catcode`\@=11

\def\marginnote#1{}
\newcount\hour
\newcount\minute
\newtoks\amorpm
\hour=\time\divide\hour by60
\minute=\time{\multiply\hour by60 \global\advance\minute by-\hour}
\edef\standardtime{{\ifnum\hour<12 \global\amorpm={am}%
             \else\global\amorpm={pm}\advance\hour by-12 \fi
             \ifnum\hour=0 \hour=12 \fi
             \number\hour:\ifnum\minute<10 0\fi\number\minute\the\amorpm}}
\edef\militarytime{\number\hour:\ifnum\minute<10 0\fi\number\minute}

\def\draftlabel#1{{\@bsphack\if@filesw {\let\thepage\relax
        \xdef\@gtempa{\write\@auxout{\string
           \newlabel{#1}{{\@currentlabel}{\thepage}}}}}\@gtempa
        \if@nobreak \ifvmode\nobreak\fi\fi\fi\@esphack}
             \gdef\@eqnlabel{#1}}
\def\@eqnlabel{}
\def\@vacuum{}
\def\draftmarginnote#1{\marginpar{\raggedright\scriptsize\tt#1}}

\def\draft{\oddsidemargin -.5truein
             \def\@oddfoot{\sl preliminary draft \hfil
             \rm\thepage\hfil\sl\today\quad\militarytime}
             \let\@evenfoot\@oddfoot \overfullrule 3pt
             \let\label=\draftlabel
             \let\marginnote=\draftmarginnote
        \def\@eqnnum{(\theequation)\rlap{\kern\marginparsep\tt\@eqnlabel}%
\global\let\@eqnlabel\@vacuum}  }

\def\preprint{\twocolumn\sloppy\flushbottom\parindent 2em
             \leftmargini 2em\leftmarginv .5em\leftmarginvi .5em
             \oddsidemargin -.5in    \evensidemargin -.5in
             \columnsep .4in \footheight 0pt
             \textwidth 10.in        \topmargin  -.4in
             \headheight 12pt \topskip .4in
             \textheight 6.9in \footskip 0pt
             \def\@oddhead{\thepage\hfil\addtocounter{page}{1}\thepage}
             \let\@evenhead\@oddhead \def\@oddfoot{} \def\@evenfoot{} }

\def\numberbysection{\@addtoreset{equation}{section}
             \def\theequation{\thesection.\arabic{equation}}}

\def\underline#1{\relax\ifmmode\@@underline#1\else
             $\@@underline{\hbox{#1}}$\relax\fi}

\catcode`@=12
\relax

\def\figcap{\section*{Figure Captions\markboth
             {FIGURECAPTIONS}{FIGURECAPTIONS}}\list
             {Figure \arabic{enumi}:\hfill}{\settowidth\labelwidth{Figure
999:}
             \leftmargin\labelwidth
             \advance\leftmargin\labelsep\usecounter{enumi}}}
 \relax
\def\tablecap{\section*{Table Captions\markboth
             {TABLECAPTIONS}{TABLECAPTIONS}}\list
             {Table \arabic{enumi}:\hfill}{\settowidth\labelwidth{Table
999:}
             \leftmargin\labelwidth
             \advance\leftmargin\labelsep\usecounter{enumi}}}
 \relax
\def\reflist{\section*{References\markboth
             {REFLIST}{REFLIST}}\list
             {[\arabic{enumi}]\hfill}{\settowidth\labelwidth{[999]}
             \leftmargin\labelwidth
             \advance\leftmargin\labelsep\usecounter{enumi}}}
 \relax
\makeatletter
\newcounter{pubctr}
\def\publist{\@ifnextchar[{\@publist}{\@@publist}}
\def\@publist[#1]{\list
             {[\arabic{pubctr}]\hfill}{\settowidth\labelwidth{[999]}
             \leftmargin\labelwidth
             \advance\leftmargin\labelsep
             \@nmbrlisttrue\def\@listctr{pubctr}
             \setcounter{pubctr}{#1}\addtocounter{pubctr}{-1}}}
\def\@@publist{\list
             {[\arabic{pubctr}]\hfill}{\settowidth\labelwidth{[999]}
             \leftmargin\labelwidth
             \advance\leftmargin\labelsep
             \@nmbrlisttrue\def\@listctr{pubctr}}}
 \relax
\makeatother
\newskip\humongous \humongous=0pt plus 1000pt minus 1000pt

\newif\ifdtup

\relax

\def\ie{{\it i.e.}}

\def\be{\begin{equation}}
\def\ee{\end{equation}}
\def\ba{\begin{eqnarray}}
\def\ea{\end{eqnarray}}

\def\diag{{\rm diag}}

\begin{document}

\begin{titlepage}
\begin{flushright}
CITUSC/02-016\\
CALT-68-2385\\
hep-th/0205184\\
\end{flushright}

\begin{center}

{\large\bf
{Matter from $G_2$ Manifolds}}\\[5mm]
{\bf Per Berglund\footnote{e-mail: berglund@citusc.usc.edu} } \\[1mm]
                           CIT-USC Center for Theoretical Physics\\
                           Department of Physics and Astronomy\\
                           University of Southern California\\
                           Los Angeles, CA 90089-0484\\[2mm]
{\bf Andreas Brandhuber\footnote{e-mail: andreas@theory.caltech.edu} } \\[1mm]
                           Department of Physics\\
                           California Institute of Technology\\
                           Pasadena, CA 91125\\[1mm]
and \\[1mm]
                            CIT-USC Center for Theoretical Physics\\
                           University of Southern California\\
                           Los Angeles, CA 90089-0484\\[5mm]

{\bf ABSTRACT}\\[3mm]
\parbox{4.9in}{We describe how chiral matter charged under
$SU(N)$ and $SO(2N)$ gauge groups arises from
codimension seven singularities in compactifications of
M-theory on manifolds with $G_2$ holonomy. The geometry of these
spaces is that of a cone over a six-dimensional Einstein space
which can be constructed by (multiple) unfolding of
hyper-K\"ahler quotient spaces.
In type IIA the corresponding picture is given by
stacks of intersecting D6-branes and chiral matter
arises from open strings stretching between them.
Usually one obtains (bi)fundamental representations but
by including orientifold six-planes in the type IIA picture
we find more exotic representations like the anti-symmetric,
which is important for the study of $SU(5)$ grand unification,
and trifundamental representations.
We also exhibit many cases where the $G_2$ metrics can be
described explicitly, although in general the metrics on
the spaces constructed via unfolding are not known.}
\end{center}
\end{titlepage}

\section{Introduction}

Compactifications of M-theory/string theory to $D=4$ dimensions with
$N=1$ supersymmetry can be obtained in a number of ways.
The historic approach, which is still very much viable, is  via the
heterotic $E_8\times E_8$ theory \cite{gross}. The compact manifold is a
Calabi-Yau three-fold and in addition a choice of vector bundle has
to be made, breaking the $E_8\times E_8$ gauge symmetry \cite{candelas}.
For example,
the standard embedding of identifying the spin connection
with the gauge connection
gives rise to $h_{2,1}$ chiral multiplets in the
${\bf 27}$ and $h_{1,1}$ chiral multiplets in the
${\bf \overline{27}}$ of $E_6$, where
$h_{2,1}$ is the number of complex structure deformations
and $h_{1,1}$ is the number of K\"ahler deformations of the Calabi-Yau
three-fold.

An alternative approach, which is dual to another heterotic
compactification \cite{bobbyed},
is to compactify M-theory  on a seven dimensional
real manifold, $X$, with $G_2$ holonomy.
In this case the duality is inherited from fiberwise application
of the duality between M-theory on a K3 manifold
and the heterotic string on
a $T^3$ with a choice of vector bundle~\cite{witten95}.
Contrary to the heterotic scenario,
compactification of M-theory on a $G_2$ manifold
does not in general lead to any charged chiral matter
nor to non-abelian gauge symmetries as long as $X$ is smooth.
Rather, one gets $b_2$ $U(1)$
vector multiplets and $b_3$ neutral chiral multiplets, where $b_q$
gives the number of $q$-cycles on $X$ \cite{pato}.
However, the situation changes drastically when $X$ admits
singularities. In particular, codimension four and seven
singularities lead to non-abelian gauge enhancement and
charged chiral matter, respectively
\cite{bobby,amv,aw1,witten,bobbyed}.\footnote{Note that this is
not a complete list of the possible singularities classified in
terms of their codimension $p$. Other interesting cases are
$p=1,5,6$, where $p=1$ corresponds to a boundary that localizes
$E_8$ gauge symmetry \cite{hw},
$p = 5$ corresponds to a singularity
of the type studied in \cite{seiberg} and
finally $p=6$ are singularities typically encountered in
Calabi-Yau three-folds which produce non-chiral matter.}
These compactifications also have an interesting interpretation
in type IIA string theory, where they correspond to compactification
on six-manifolds with RR two-form fluxes and/or D6-branes/$\mathcal{O}6$
planes wrapped on supersymmetric three-cycles \cite{jaume,kachru}.
For examples of such type IIA compactifications which
give rise to chiral fermions and which admit lifts to M-theory
compactifications on $G_2$ manifolds, see \cite{cvetic}.

In this paper, we continue the work of Acharya and
Witten~\cite{bobbyed}, who showed
how chiral matter charged under unitary gauge groups
appear at singularities of $G_2$
manifolds. These manifolds are constructed as
circle quotients of conical hyper-K\"ahler (HK) eight-manifolds.
In particular, we generalize their work to the antisymmetric
representation of $SU(n)$ and the fundamental of $SO(2n)$.
The former in particular, is essential in attempts of
constructing realistic models of grand unified theories such as
$SU(5)$ from string/M-theory.
We describe how the corresponding M-theory backgrounds can be
reduced to type IIA backgrounds that consist of
intersecting D6-branes/$\mathcal{O}6$ planes.
This straightforwardly generalizes
a collection of two and three stacks of intersecting
D6-branes to $k$ groups of $n_i\,,i=1,\ldots,k$ D6-branes (see
also~\cite{gutong}).
For $k=3$ we compare our results with those of~\cite{bobbyed} which
have an M-theory description as $\mathbb{R}^3$-bundles over
$\mathbb{WCP}^2_{n_1,n_2,n_3}$ if at least two of the indices $n_i$ are
equal. In the presence of an $\mathcal{O}6$ plane the M-theory
background is a $\mathbb{Z}_2$ orbifold thereof.
We also describe how trifundamental matter charged under
$SU(3) \times SU(3) \times SU(2)$ and $SU(4) \times SU(3) \times
SU(2)$ are obtained from unfolding HK quotients~\cite{kronheimer} of
$E_6$ and $E_7$ singularities, respectively.

The rest of the paper is organized as follows: in section 2 we
generalize the construction of $G_2$ manifolds by multiple
unfoldings of HK quotient singularities.
In section 3 we show that the space arising from double unfoldings
under certain conditions
can be mapped to a different construction of $G_2$ manifolds as cones over
twistor spaces, which makes it possible to find the
metrics explicitly. In section 4 we include orientifolds in the type IIA
picture which in M-theory lifts to $\mathbb{Z}_2$ orbifolds of manifolds
constructed
via multiple unfoldings. This gives rise to chiral multiplets in the
anti-symmetric representation of $SU(n)$ gauge groups in addition to
multiplets in the bifundamental representation.
We also extend our work to exceptional gauge groups, which give rise to
matter charged under more than two non-abelian gauge groups for which
there is no perturbative type IIA description available.
Finally, we end with some comments and discussions in section 5 while details
of the map from M-theory to type IIA can be found in appendix A and
the realizations of $D_n$ and $E_n$ singularities are given in
appendices B and C, respectively.

\section{Unfolding hyper-K\"ahler quotients}

In this section we review and generalize a particular construction of
$G_2$ manifolds introduced by Acharya and Witten in \cite{bobbyed}.
The manifolds are $U(1)$ quotients
of HK manifolds where the $U(1)$
is chosen such that it commutes with the $SU(2)$ symmetry of the HK
manifold which permutes the three complex structures, and can be
obtained by {\it unfolding} four-dimensional HK quotient singularities.
This unfolding, which will be described in more detail later, is
nothing but a fibration of an ADE singularity over a three-dimensional
base, $B$. If there are no singularities worse than the codimension
four ADE singularity we would just obtain an $N=1$ vector multiplet
with ADE gauge group and $b_1(B)$ chiral multiplets in the adjoint.
But in the construction of \cite{bobbyed} the base manifold $B$ has
one special point (or possibly several) over which the singularity
is enhanced e.g. $A_{n-1} \to A_{n}$. In the dual heterotic string
theory this worsening of the singularity is reflected by a jump in
the rank of the gauge bundle at the same point(s)
in the base $B$ as in M-theory. (We refer the
reader to \cite{bobbyed} for more details on the heterotic picture.)
In the case of a symmetry enhancement $A_{n-1} \to A_n$
the gauge theory is $SU(n) \times U(1)$ SYM with chiral matter
in the ${\bf n}_{-1}$ representation\footnote{In the following
we label representations of non-abelian group factors by their
dimensionality. Hence, the fundamental representation of $SU(n)$ is
denoted by ${\bf n}$ and its complex conjugate by $\overline{\bf n}$.
A charge with respect to an abelian factor is denoted by a subscript, e.g.
${\bf n}_{Q}$.}.
The $SU(n)$ vector multiplet is produced by a co-dimension four
$A_{n-1}$ singularity, whereas the $U(1)$ comes from conventional
Kaluza-Klein reduction \cite{aw1}.
As was shown in \cite{witten} the anomaly
from the chiral multiplet is canceled by an anomaly inflow
mechanism from the bulk~\cite{ghm}. This is the same mechanism that is
at work to cancel the anomaly of chiral matter that arises from
open strings stretched between configurations of intersecting
D-branes which preserve four supercharges \cite{bdl}.
Of course, this is more than a formal analogy since the $G_2$
singularities have an interpretation in type IIA string theory
as sets of D6-branes intersecting over flat $\mathbb{R}^6$ \cite{aw1}
as we will show in this section (with more details in appendix A).

Interestingly, there exists a closely related construction in Type IIA
of Calabi-Yau threefold singularities which give rise to charged
matter \cite{katzvafa}. The threefold is fibered by K3's over a complex
one-dimensional base $Q$ in a manner that over generic points of $Q$
the singularity is $G$ but over one point the symmetry is enhanced
$G \to G^\prime$. There exists a fruitful interplay between these two
constructions and we will make this more explicit in examples
presented later in this section.

We will start by reviewing Kronheimers HK quotient construction of
ADE singularities~\cite{kronheimer}
which we will use to obtain charged chiral matter (see also
\cite{bobbyed}). The starting point is an $A_{n}$ singularity which
is locally equivalent to $\mathbb{R}^4/\mathbb{Z}_{n+1}$.
The basic idea is that $\mathbb{R}^4/\mathbb{Z}_{n+1}$ can be realized
as the vacuum manifold of a particular linear sigma model (LSM) and
will be denoted as $\mathbb{H}^{n+1}//K$, where $K$ is the LSM gauge group
which in this case is $U(1)^{n}$ and each factor of $\mathbb{H}$ denotes
the four scalars of a hypermultiplet $\Phi_i ~,~ i = 0, \ldots n$.
The charges of the hypermultiplets follow from the affine Dynkin
diagram of $A_n$, also denoted as the quiver diagram of the LSM.
The hypermultiplet $\Phi_i$ has charge
$+1$ and $\Phi_{i-1}$ has charge $-1$
under the $i^\mathrm{th}$ $U(1)$ and all other charges are zero.
The vacuum manifold, denoted in short by $\mathbb{H}^{n+1}//K$, of this theory
is obtained by imposing the
D/F-term constraints and dividing by the gauge group $K$.
This manifold has real
dimension $4 (n+1) - 3 n - n = 4$.
The $3 n$ D/F-term constraints form $n$ triplets under the $SU(2)_\mathcal{R}$
symmetry of the LSM and can be expressed as linear combinations
of the moment map $\mu : \mathbb{H} \to \mathbb{R}^3$:
\be
\Phi = (M, \overline{M}) \to \left(
\begin{array}{c}
\Re M \overline{M} \\
\Im M \overline{M} \\
{M}^\ast M - {\overline{M}}^\ast \overline{M}
\end{array} \right) \equiv (\Phi, \vec{\sigma} \Phi) ~,
\label{mommap}
\ee
where $M$ and $\overline{M}$ are two complex scalars contained in
$\Phi$, and $\sigma_i$
are the standard hermitian Pauli matrices. Note that $\overline{M}$ has
opposite charge of $M$ and together they form a doublet of $SU(2)_\mathcal{R}$.
Therefore, and because of the charge assignments of the scalars, the D/F-term
of the $l$-th $U(1)$ is
\be
-(\Phi_{l-1}, \vec{\sigma} \Phi_{l-1})+(\Phi_l, \vec{\sigma} \Phi_l)
       = \vec{t}_l ~,
\label{mmap2}
\ee
where $\vec{t}_l$ is a triplet of FI-parameters.

The $A_n$ singularity in the form $\mathbb{R}^4/\mathbb{Z}_{n+1}$ is
recovered when we set all FI-parameters $\vec{t}_l = 0$.
The quotient by $K$ is implemented by introducing gauge invariant
meson and baryon like combinations of the complex scalars
\be
x{=}\prod M_i = z_1^{n+1}\,,\,y{=}\prod \overline{M}_i {=} z_2^{n+1}\,,\,
       z{=}M_0 \overline{M}_0 {=}\ldots {=}  M_n \overline{M}_n {=} z_1 z_2~,
\label{baryon}
\ee
which obey
the standard equation for an $A_n$ singularity $x y = z^{n+1}$.
On the other hand when some or all of the moment maps are non-zero
we get partial resolutions of the singularity in which case either
the K\"ahler parameter of certain shrunk two-cycles is blown up
and/or the complex structure equations is deformed to
$x y = z^{n+1} - u_2 z^{n-1} - u_3 z^{n-2} - \ldots - u_{n+1}$.

This is the starting point for Katz and Vafa's construction of charged
matter from geometry \cite{katzvafa}.
The key idea is to deform a higher singularity
into a lower by complex structure deformations which vary over  space.
This amounts to replacing the constant deformation parameters $u_i$
by functions over a base which we take to be $\mathbb{CP}^1$ and is
parametrized by $t$. The breaking of an $A_r \to A_{r-1}$ singularity
is given by the deformation
$x y = z (z + t)^r$ which gives rise to matter hypermultiplets
in the ${\bf r}_{(-1)}$ representation of $SU(r) \times U(1)$.
This can be readily generalized
in two ways. We can reduce the rank by one in more than one fashion, e.g.
$x y = z^{n} (z+t)^{m}$ which corresponds to
$A_{n+m-1} \to A_{n-1} \times A_{m-1}$ and gives matter in the
${\bf (n,\overline{m})}$ of $SU(n) \times SU(m)$.
The other possibility is to break the higher singularity into more
than two smaller singularities by
$x y = \prod_{i=1}^m (z + c_i t)^{n_i}$, where $c_i$ are
arbitrary constants, which corresponds to
$A_{N-1} \to \prod A_{n_i-1}$ with $N = \sum_{i=1}^m n_i$.
In the context of $G_2$ manifolds we will soon employ a similar
construction which we call $m \!\! - \!\! 1$-fold
unfolding of the singularity.

The construction shortly reviewed in the last two paragraphs \cite{katzvafa}
was mainly considered in the context of Calabi-Yau compactifications of
type IIA string theory
where it leads to charged hypermultiplets. In \cite{bobbyed} a closely
related construction of singularities of seven-dimensional spaces
with $G_2$ holonomy was presented. These spaces are to be seen in the
context of $M$-theory compactification to four dimensions with $\mathcal{N}=1$
supersymmetry and charged chiral matter.
As before one is interested in fibering an $ADE$ singularity over
a base manifold such that the singularity is maximal over a special point
but smaller at generic points. However, the base manifold must be
three-dimensional and the space cannot be written simply as
a complex structure deformation of an $ADE$ singularity.
As described in \cite{bobbyed} this is achieved by unfolding a HK quotient
singularity $\mathbb{H}^{n+1}//K$ by setting all moment maps to zero
except for one
\be
-(\Phi_{i-1}, \vec{\sigma} \Phi_{i-1})+(\Phi_i, \vec{\sigma} \Phi_i)=0
\label{notzero}
\ee
with $i = 1, \ldots, k-1, k+1, \ldots, n$ where $k$
has to be omitted. The corresponding unconstrained moment map is
\be
-(\Phi_{k-1}, \sigma \Phi_{k-1})+(\Phi_k, \sigma \Phi_k)= \vec{t}_1 ~,
\label{notcnstrnt}
\ee
where the three-vector $\vec{t}_1$ should be thought of as
parametrizing the three-dimensional base manifold. This clearly
corresponds to a seven-manifold, which is conical and non-compact,
and the symmetry breaking is of the
form $A_{n-1} \to A_{k-1} \times A_{n-k-1}$. This is the prototype example
of an unfolded $A_{n-1}$ singularity and in this case the manifold was shown
to be a cone over $\mathbb{WCP}_{k,k,l,l}$ with $l = n-k$~\cite{bobbyed}.
The chiral matter localized at the singularity is in the representation
${\bf (k,\overline{n-k})}$.

Let us explain in some more detail how the generalization
to double unfoldings works\footnote{It is straightforward
to generalize our construction to multiple unfoldings, and we will return to
the general case at the end of this section. However, in this paper we will
be mainly concerned with double unfoldings.}.
We start with an $A_{N-1}$ singularity realized
as an HK quotient, where $N = n_1+n_2+n_3$.
Let two of the moment maps be unrestricted
\ba
&& -(\Phi_{n_1-1}, \vec{\sigma} \Phi_{n_1-1}) +
         (\Phi_{n_1}, \vec{\sigma} \Phi_{n_1})= \vec{t}_1 ~,~ \nonumber \\
&& -(\Phi_{n_1 + n_2-1}, \vec{\sigma} \Phi_{n_1+n_2-1})+
         (\Phi_{n_1+n_2}, \vec{\sigma} \Phi_{n_1+n_2})= \vec{t}_2 ~,
\label{unrest}
\ea
and otherwise
\be
(\Phi_{i-1}, \vec{\sigma} \Phi_{i-1}) = (\Phi_i, \vec{\sigma} \Phi_i) ~,
\label{restr}
\ee
with $i=1, \ldots, \widehat{n}_1, \ldots, \widehat{n_1+n_2}, \ldots
,N-1$.

As it stands this corresponds to a ten-dimensional manifold since
this is a fibration over a six-dimensional space spanned by $\vec{t}_{1,2}$.
In a moment we will restrict to a three-dimensional subspace via a
linear constraint between the two three-vectors. But let us first
rephrase the D/F-term equations in a way that takes care of the
gauge symmetry we have to divide by. As usual this is done by
introducing gauge invariant combinations of the scalar fields
$\Phi_i = (M_i,\overline{M}_i)$. We perform a double unfolding
in which we leave the D/F-terms corresponding to
the $n_1$-th and $(n_1+n_2)$-th $U(1)$
unconstrained,
\ba
&&\overline{M}_i M_i = \overline{M}_{i+1} M_{i+1} ~,~
|M_i|^2 - |\overline{M}_i|^2 = |M_{i+1}|^2 - |\overline{M}_{i+1}|^2 ~,
        \nonumber \\
&&\overline{M}_{n_1-1} M_{n_1-1} - \overline{M}_{n_1} M_{n_1} =
t_1^{(1)} + i t_1^{(2)} \equiv c_1 ~,
\nonumber \\
&&|M_{n_1-1}|^2 - |\overline{M}_{n_1-1}|^2 - |M_{n_1}|^2 +
|\overline{M}_{n_1}|^2 = t_1^{(3)} \equiv r_1 ~, \nonumber \\
&&\overline{M}_{n_1+n_2-1} M_{n_1+n_2-1} -
        \overline{M}_{n_1+n_2} M_{n_1+n_2} = t_2^{(1)} + i t_2^{(2)}
\equiv c_2 ~, \nonumber \\
&&|M_{n_1+n_2-1}|^2 - |\overline{M}_{n_1+n_2-1}|^2 - |M_{n_1+n_2}|^2 +
|\overline{M}_{n_1+n_2}|^2 = t_2^{(3)} \equiv r_2 ~,
\label{dblunfold}
\ea
with $i=0,\ldots ,n_1-2 , n_1,\ldots ,n_1+n_2-2, n_1+n_2,\ldots ,
N-2$.
This corresponds to the breaking
pattern $A_{N-1} \to A_{n_1-1} \times A_{n_2-1} \times A_{n_3-1}$.
Eqs.~(\ref{dblunfold}) can be solved by introducing $K'$ invariant fields
where $K = K' \times U(1)_{n_1} \times U(1)_{n_1+n_2}$.
These baryonic fields are given by
\be
\begin{array}{lcl}
       \prod_{i=0}^{n_1-1} M_i = z_1^{n_1} &~,~&
\prod_{i=0}^{n_1-1} \overline{M}_i = z_2^{n_1} ~, \\[3mm]
       \prod_{i=n_1}^{n_1+n_2-1} M_i = z_3^{n_2} &~,~&
\prod_{i=n_1}^{n_1+n_2-1} \overline{M}_i = z_4^{n_2} ~, \\[3mm]
       \prod_{i=n_1+n_2}^{N-1} M_i = z_5^{n_3}& ~,~&
\prod_{i=n_1+n_2}^{N-1} \overline{M}_i = z_6^{n_3} ~,
\label{doublebaryon}
\end{array}
\ee
where the $z_i$ are neutral under $K'$ and carry charges
\be
\begin{array}{lcl}
U(1)_{n_1} & : & (-1/n_1,1/n_1,1/n_2,-1/n_2,0,0) ~, \nonumber \\
U(1)_{n_1+n_2} & : & (0,0,1/n_2,-1/n_2,-1/n_3,1/n_3) ~,
\label{u1action}
\end{array}
\ee
under the remaining gauge symmetries.

In analogy with~(\ref{baryon}) we can introduce a different set
of meson and baryon like variables.
\ba
&&    x = \prod_{i=0}^{N-1} M_i ~,~
        y = \prod_{i=0}^{N-1} \overline{M}_i ~, \nonumber \\
&&    z^{n_1} = \prod_{i=0}^{n_1-1} M_i \overline{M}_i ~,~
        w^{n_2} = \prod_{i=n_1}^{n_1+n_2-1} M_i \overline{M}_i ~,~
        u^{n_3} = \prod_{i=n_1+n_2}^{N-1} M_i \overline{M}_i ~,
\label{barymes}
\ea
with $N=n_1+n_2+n_3$. They obey the following 
equation
\be
x y = z^{n_1} w^{n_2} u^{n_3} = z^{n_1} (z - c_1)^{n_2}
(z - c_1 - c_2)^{n_3}
\label{hysurf}
\ee
where in the last equality we have used the F-term equations from the
$n_1$-th and $n_1+n_2$-th $U(1)$ (see Eq.~(\ref{dblunfold})).
This allows us to make contact with the generalized construction
of Katz and Vafa~\cite{katzvafa}.
When $c_1=c_2=0$ Eq.~(\ref{hysurf}) describes an
$A_{N-1}$ singularity. For $c_1 \neq 0$ and
$c_1+c_2 \neq c_1$ we have $A_{N-1} \to A_{n_1-1} \times
A_{n_2-1} \times A_{n_3-1}$.

Let us study the geometry of the (singular) non-compact manifold given by the
above HK quotient construction. The ambient space is
$\mathbb{C}^6$ subject to the $U(1)^2$ actions given by (\ref{u1action}).
In order for this
to describe a seven-dimensional manifold
we impose a constraint of the type
\be
\vec{t}_1 + R \cdot \vec{t}_2 = \vec{c}
\label{linconst}
\ee
where the $\vec{t}_i$ are the relevant moment maps from~(\ref{dblunfold})
\ba
&& z_1 z_2 - z_3 z_4 = t^{(1)}_1 + i t^{(2)}_1 \equiv c_1 ~, \nonumber \\
&& (|z_1|^2 - |z_2|^2) - (|z_3|^2 - |z_4|^2) = t^{(3)}_1 \equiv r_1
~, \nonumber \\
&& z_3 z_4 - z_5 z_6 = t^{(1)}_2 + i t^{(2)}_2 \equiv c_2 ~, \nonumber \\
&& (|z_3|^2 - |z_4|^2) - (|z_5|^2 - |z_6|^2) = t^{(3)}_2 \equiv r_2 ~.
\label{hyper1}
\ea
Non-vanishing $\vec{c}$ in Eq.~(\ref{linconst}) gives rise to
interesting geometrical transitions, but
we are mainly interested in conical $G_2$ holonomy spaces,
that give rise to chiral fermions, and so we set $\vec{c} = 0$.
It is then easy to see that the set of equations (\ref{hyper1}) is
invariant under $z_i \to \lambda z_i$ with $\lambda \in \mathbb{R}^+$.
Hence the space is a cone over a six dimensional base, $Y$.
Furthermore, by enlarging the $U(1)^2$ actions on $\mathbb{C}^6$ to
$(\mathbb{C}^*)^2$ we have a toric variety, $M$, of
$dim_\mathbb{C}=4$~\footnote{Alternatively, we can use the equivalent
formulation of a toric variety in terms of the symplectic
quotient construction in which each $U(1)$ action is combined with the
D-term component of the corresponding  moment map, see e.g.
\cite{witten93}, \cite{mp}.}.
Because of the linear constraint~(\ref{linconst}) the complex part
of~(\ref{hyper1})
gives rise to a quadratic relation between the $z_i$. Thus, the base
$Y$ is obtained as a hypersurface in $M$. It is interesting to note that
although the tools of algebraic geometry are not directly available to
describe the geometry of the $G_2$ manifold itself, in the case of
a conical geometry the base carries a K\"ahler structure and it often
has singularities in codimension four (see appendix A).
To be more precise, the $G_2$ holonomy condition implies that
the base is a nearly K\"ahler Einstein manifold \cite{gray}.

It is instructive to analyze what these manifolds correspond to
when we reduce to Type IIA string theory along a circle
that is  generated by a $U(1)$ action with charges
$(1,-1,0,0,0,0)$ on the variables $z_i$. We leave the derivation
to appendix A and only state the results.
In the case of double unfolding we find that the Type IIA
background describes the intersection of three stacks of D6
branes at supersymmetric angles over $\mathbb{R}^6$, where the
number of the D6-branes in the $i$-th stack is given by three integers
$n_i$. This allows us to derive the expected chiral
matter directly and we find multiplets in the representation
${\bf (n_1,\overline{n}_2,1) + (1,n_2,\overline{n}_3) +
(\overline{n}_1,1,n_3)}$ of $SU(n_1) \times SU(n_2) \times SU(n_3)$.
Hence, the physics at these singularities is quite
straightforward to understand, unlike the case of $G_2$
singularities from twistor spaces \cite{bobbyed}
which are in general harder to analyze and
probably contain new, interesting physics. (For recent work,
see e.g.~\cite{calin,klaus}).
However, in section 3 we show that certain singularities
constructed by double unfolding can be related to a particular
class of twistor space constructions, and, therefore, the $G_2$
metric is known explicitly in those examples.

It is natural to generalize this construction by considering
multiple unfoldings of $A_N$ singularities by leaving $m$ D/F
terms unconstrained. In order to obtain an appropriate
seven-manifold we have to introduce $m \! - \! 1$ linear relations
of the type~(\ref{linconst}). By using a generalization of the
discussion in appendix A we find that this space corresponds
to an intersection of $m+1$ stacks of D6-branes with
multiplicities $n_i$. In this case we obtain matter in
the bifundamental representations of
$SU(n_1) \times \ldots \times SU(n_{m+1})$
charged under all possible
combinations of two gauge group factors which in total give
$m (m+1) \over 2$ chiral multiplets.

\section{Relation between unfolding and twistor space
construction of $G_2$ singularities}

In this section we will show how the procedure presented in
section 2, specialized to a double unfolding of an $A_n$
singularity, can be related to the twistor space construction
of $G_2$ holonomy cones considered in \cite{bobbyed}.
In particular we will find a connection with twistor spaces
over weighted projective spaces $\mathbb{WCP}^2_{k_1,k_2,k_3}$.
The advantage of the twistor space constructions is that they
automatically provide a $G_2$ holonomy metric without extra work.
However, the physics is harder to understand \cite{bobbyed} and an
interpretation in terms of sets of D6-branes intersecting over
$\mathbb{R}^6$ is not always available \cite{aw1}. As was explained in the
previous section the unfolding procedure always gives singularities
that can be viewed as intersecting D6-branes in Type IIA string
theory and their physical interpretation is more accessible via
duality with heterotic string theory. Unfortunately, it is very
hard in general to find the $G_2$ holonomy metrics explicitly.
But in order to understand the physics this is not really needed
and duality with Type IIA strings guarantees that intersections
of D6-branes at supersymmetric angles have lifts to $G_2$ holonomy
metrics in M-theory.

Let us quickly review the example studied in \cite{bobbyed}
pertaining to a cone on the twistor space over
$\mathbb{WCP}^2_{k_1,k_2,k_3}$.
For this purpose we consider the vacuum manifold of a $U(1)$ gauge
theory coupled to three hypermultiplets which contain complex
scalars $a_i, \bar{b}_i, i = 1, 2, 3$ with $U(1)$ charges
$(q_i, -q_i)$. The vacuum manifold is given by the D-term and F-term
constraints modulo gauge transformations with the triplet of
FI-terms set to zero. After rescaling the scalars,
$x_i = \sqrt{q_i} a_i$ and $y_i = \sqrt{q_i} b_i$, we find
the equations
\be\label{su3cone}
\sum_{i=1}^3 |x_i|^2 - |y_i|^2 = 0 ~,~\sum_{i=1}^3 x_i \bar{y}_i = 0 ~.
\ee
These equations describe a cone over $SU(3)$ embedded in
$\mathbb{C}^6 = \mathbb{H}^3$. The conical structure follows
from the invariance of the equations (\ref{su3cone}) under
the simultaneous rescaling $x_i \to \lambda x_i$ and $y_i \to \lambda y_i$.
In order to find the base of the cone we fix this invariance by
intersecting the cone with an eleven-sphere
$\sum |x_i|^2 + |y_i|^2 = 2 \subset \mathbb{C}^6$.
Together with (\ref{su3cone}) we find
\be\label{ortnor}
\sum_{i=1}^3 |x_i|^2 = |y_i|^2 = 1 ~,~\sum_{i=1}^3 x_i \bar{y}_i = 0 ~,
\ee
which are nothing but the orthonormality conditions of two complex
three-vectors $\vec{x}$ and $\vec{y}$. Introducing a third vector
$\vec{z} = \overline{\vec{x} \times \vec{y}}$
we can package these
three vectors into an $SU(3)$ matrix $M = \{ \vec{x}, \vec{y}, \vec{z} \}$.
The charges of the components of $M$ under the $U(1)$ gauge symmetry
can be summarized in a charge matrix
\be\label{qmatrix1}
\left(
\begin{array}{ccc}
q_1 & q_1 & -(q_2+q_3) \\
q_2 & q_2 & -(q_1+q_3) \\
q_3 & q_3 & -(q_1+q_2)
\end{array}
\right) ~.
\ee
Furthermore, the $SU(3)$ manifold admits an $SU(2)$ action that rotates
$\vec{x}$ and $\vec{y}$ into each other but leaves $\vec{z}$ fixed.
It acts on $M$ by right multiplication
\be\label{su2action}
M \to M \cdot
\left(
\begin{array}{cc}
A_{SU(2)} & 0 \\
0 & 1
\end{array}
\right) ~.
\ee
This $SU(2)$ contains an abelian subgroup $U(1)_H$ that acts by
right multiplication with $\diag \{e^{i \alpha}, e^{-i \alpha}, 1\}$
and the corresponding charge table is
\be\label{qmatrix2}
\left(
\begin{array}{ccc}
1 & -1 & 0 \\
1 & -1 & 0 \\
1 & -1 & 0
\end{array}
\right) ~.
\ee
The identification of the twistor space proceeds as follows.
The base of the twistor space is given by $SU(3)/(SU(2)\times U(1))$.
First note that the quotient $SU(3)/SU(2)$ is simply a copy
of $S^5$ since it is given by $\vec{z}$ which is $SU(2)$
invariant and obeys $|\vec{z}|^2=1$. We have to further divide
by $U(1)$ where the charges of the $z_i$ can be read off from
the third column of (\ref{qmatrix1}).
By a redefinition of the $U(1)$ generator we can remove
the overall minus sign and find that $S^5/U(1)$ is
$\mathbb{WCP}^2_{k_1,k_2,k_3}$ with $k_i = \{q_2+q_3,q_1+q_3,q_1+q_2\}$, or,
if all $q_i$ are odd $k_i = \frac{1}{2}\{q_2+q_3,q_1+q_3,q_1+q_2\}$.

Finally, the $\mathbb{S}^2$ fiber of the twistor space is given by
$SU(2)/U(1)_H$, and the twistor space is given by
$SU(3)/(U(1)\times U(1)_H)$, where $U(1)_H$ is a
subgroup of the $SU(2)$ group action (\ref{su2action}).
The $G_2$ holonomy manifold constructed as the cone over
the twistor space has a codimension seven singularity.
Since in the case at hand the base of the cone is homogeneous
$G/H$ a resolution of the singularity may be provided
by a cohomogeneity one metric of the type
$ds^2 = dr^2 + g_{G/H}(r)$, where $g_{G/H}(r)$ is an $r$
dependent metric on the so-called principal orbit $G/H$
and $r \geq r_0$. The necessary conditions for a
smooth resolution \cite{cleyton} are the existence
of a singular orbit $G/K$ with $G \supset K \supset H$
which has finite volume at $r=r_0$ and that $K/H$ becomes
a round sphere $S^n$. This means that the cone is resolved to
a $\mathbb{R}^{n+1}$ bundle over $G/K$.
In our case $K = SU(2) \times U(1)$ is the only possibility
with $G/K = \mathbb{WCP}^2_{k_1,k_2,k_3}$ and $K/H = \mathbb{S}^2$.
Note, that in general this only removes the codimension seven
singularity but $G/K$ itself may have orbifold singularities, as
is the case for weighted projective spaces \cite{galicki}.

In general, a twistor space can be constructed over any four-manifold
$M$ which is Einstein with positive curvature and has a
self-dual Weyl tensor. In that case we can immediately
write down a $G_2$ holonomy metric \cite{bryant,gibbons}
\be\label{g2metric}
ds^2 = \frac{dr^2}{1-r_0^4/r^4} + \frac{r^2}{2} ds^2_M +
\frac{r^2}{4} \left(1-r_0^4/r^4\right) |D t|^2 ~,
\ee
with
$D t_i = d t_i + \epsilon_{ijk} A^j t^k ~,~ \sum t_i^2 = 1$.
The $A^i$ are three one-form gauge fields with anti-selfdual
field strengths $F^i = dA^i + \frac{1}{2} \epsilon_{ijk} A^j \wedge A^k$.
The parameter $r_0$ in the metric (\ref{g2metric}) is a blow-up
parameter. For $r_0 = 0$ the metric has a conical singularity
at $r=0$, but for $r_0 > 0$ the four-manifold $M$ is blown-up to finite
size in the interior $r=r_0$ and the total space is an
$\mathbb{R}^3$ bundle over $M$.
However, the manifold $M$ itself can have
orbifold singularities as is the case for weighted projective spaces.


Now we wish to relate this approach to the kind of models constructed
by double unfoldings of HK quotient spaces as introduced in section 2.
The two non-zero moment maps correspond
to the gauge factors $U(1)_{n_1} \equiv H_1$ and $U(1)_{n_1+n_2} \equiv H_2$.
The double unfolding procedure
produces  hypersurface equations in $\mathbb{C}^6$
whose explicit form can be found in (\ref{hyper1}) modulo the
$U(1)$ actions (\ref{u1action}).
This constitutes a ten-dimensional manifold and we have to impose
linear relations (\ref{linconst}) between the constants $c_i$, $r_i$
in order to obtain a seven-dimensional space.
The appropriate choice is $r_1 = r_2 ~,~ c_1 = c_2$, so that (\ref{hyper1})
becomes
\ba\label{mom3}
&& |z_1|^2 - |z_2|^2 - 2 |z_3|^2 +2 |z_4|^2 + |z_5|^2 - |z_6| = 0 ~,
\nonumber \\
&&  z_1 z_2 - 2 z_3 z_4 + z_5 z_6 = 0 ~.
\ea
where the $z_i$ have the following charges
\be
\begin{array}{c|c|c|c|c|c|c|}
       & z_1 & z_2 & z_3 & z_4 & z_5 & z_6 \\ \hline
H_1 & 1/n_1 & -1/n_1 & -1/n_2 & 1/n_2 & 0 & 0 \\ \hline
H_2 & 0 & 0 & 1/n_2 & -1/n_2 & -1/n_3 & 1/n_3 \\ \hline
\end{array}
\ee

In order to bring (\ref{mom3}) to a more familiar form,
we make the replacements
\be
(z_1, \ldots, z_6) \to (a_1, \bar{b}_1, \bar{b}_2/\sqrt{2}, -a_2/\sqrt{2},
a_3, \bar{b}_3)
\ee
which turns (\ref{mom3}) into
\be
\sum_{i=1}^3 |a_i|^2 - |b_i|^2 = 0 ~,~ \sum_{i=1}^3 a_i \bar{b}_i = 0 ~.
\ee
At this point we have achieved part of our goal
of identifying the unfolding and  twistor space constructions.
It remains to match the charges under $H_1$ and $H_2$ with
the charges in (\ref{qmatrix1}) and (\ref{qmatrix2})
under $U(1)$ and $U(1)_H$. We are of course allowed to
take arbitrary linear combinations of the generators.
After rescaling the generators of $H_{1,2}$ we can turn
the fractional charges
of the $a_i$ and $b_i$ into integers
\be
\begin{array}{c|c|c|c|c|c|c|}
       & a_1 & b_1 & a_2 & b_2 & a_3 & b_3 \\ \hline
H_1 & n_2 & n_2 & -n_1 & -n_1 & 0 & 0 \\ \hline
H_2 & 0 & 0 & n_3 & n_3 & -n_2 & -n_2 \\ \hline
\end{array}
\ee
As before we have a cone over $SU(3)$ modded out by $U(1)^2$
generated by $H_{1,2}$. By introducing the vector
$\vec{c} = \overline{\vec{a} \times \vec{b}}$ we
get the $SU(3)$ matrix
\be
\left(
\begin{array}{ccc}
a_1 & a_2 & a_3 \\
b_1 & b_2 & b_3 \\
c_1 & c_2 & c_3
\end{array}
\right) ~.
\ee
The charges of the fields in this matrix under $H_1$ and $H_2$
respectively are
\be
Q_1=\left(
\begin{array}{ccc}
n_2 & -n_1 & 0 \\
n_2 & -n_1 & 0 \\
n_1 & -n_2 & n_1-n_2
\end{array}
\right)  ~~ \mathrm{and} ~~
Q_2=\left(
\begin{array}{ccc}
0 & n_3 & -n_2 \\
0 & n_3 & -n_2 \\
n_2-n_3 & n_2 & -n_3
\end{array}
\right)
\label{qmatrixunfld}
\ee

In the simplest case we take $n_1=n_2=n_3=1$. We easily see that
$Q_1$ and $Q_1 - Q_2$ agree with the charges of $U(1)$ and $U(1)_H$
respectively.
This corresponds to the case of the cone over the twistor space of
$\mathbb{CP}^2$.

Next we take two of the $n_i$ equal. Without loss of generality we
can choose $n_1=n_2=p$ and $n_3=q$. Note that the third column
of $Q_1$ is zero which is necessary to make contact with the twistor space
construction. To complete the identification we introduce
$\tilde Q_1=\frac{1}{p} Q_1$ and $\tilde Q_2=\frac{q}{2 p} Q_1 + Q_2$
which give
\be
\tilde{Q}_1=\left(
\begin{array}{ccc}
1 & -1 & 0 \\
1 & -1 & 0 \\
1 & -1 & 0
\end{array}
\right)  ~~ \mathrm{and} ~~
\tilde{Q}_2=\left(
\begin{array}{ccc}
q/2 & q/2 & -p \\
q/2 & q/2 & -p \\
p-q/2 & p-q/2 & -q
\end{array}
\right)~.
\label{qmatrixppq}
\ee
From~(\ref{qmatrix2}) and~(\ref{qmatrix1}) this gives the twistor
space over $\mathbb{WCP}^2_{p,p,q}$.
According to \cite{bobbyed} this singularity has an interpretation
in Type IIA as an intersection of three groups of D6-branes with
multiplicities $p$, $p$ and $q$. The $N=1$ supersymmetric gauge
theory on these branes has gauge group $SU(p) \times SU(p) \times SU(q)$
and is coupled to chiral multiplets in the
$\bf{(p,\overline{p},1) + (\overline{p},1,q) + (1,p,\overline{q})}$
representation.

Finally, when all $n_i$ are distinct it is impossible to
match the two constructions as can seen from the charge
table~(\ref{qmatrixunfld}).
It turns out that this has an interesting physical interpretation.
As we explained in section 2 the unfolding procedure naturally
gives configurations of intersecting D6-branes in flat space, but it
does not provide an easy way to construct the $G_2$ holonomy metric.
On the other hand the manifolds obtained via the twistor space construction
inherit a natural $G_2$ structure and the metric can be found explicitly
\cite{bryant,gibbons}.
However, the physics of these singularities is in general much harder
to understand because of the occurrence of
codimension six singularities in addition to the much better understood
codimension four singularities.
In particular the cases based on twistor spaces of
$\mathbb{WCP}^2_{k_1,k_2,k_3}$ with all $k_i$ different does not have
a simple interpretation as an intersection of flat D6-branes over
$\mathbb{R}^6$. Only if two or three of the $k_i$ are equal is such
an interpretation available which is precisely when we can make a connection
with the unfolding construction.

\section{$D_n$  and $E_n$ hyperk\"ahler quotients} 
M-theory on ALE-spaces describing $D_n$ and $E_n$ singularities give rise to
$SO(2n)$ and $E_n$ gauge groups, respectively. (We will focus on simply laced
groups--for more details on the other, disconnected components of the space
of seven-dimensional theories with sixteen supercharges, see~\cite{deboer}.)
The enhanced gauge symmetry
comes from M2-branes wrapping vanishing 2-cycles which intersect according
to the Cartan matrix for the corresponding Lie group~\cite{witten95}.
In this section we will consider M-theory realizations of $SO(2n)$ and
$E_n$ gauge symmetries, as well as charged chiral matter, in terms of
the hyperk\"ahler quotient construction described in section~2.
We will concentrate on the HK quotient manifolds describing
the unfolding of $D_n$ singularities, and return to the
exceptional groups at the end of the section.

The procedure is carried out in two steps. First, we have to describe
the $D_n$ singularity in terms of a hyperk\"ahler quotient construction.
This is done by considering a $\mathbb{Z}_2$ orbifold of an
$A_{n'}$-model. Second, the matter is described by unfolding the
$D_n$ theory. Just as in the $N=2$ compactification of type IIA
theory on a K3-fibered Calabi-Yau manifold there are several
possibilities for how this can be done~\cite{katzvafa}, e.g. we get a
${\bf 2(n-1)}$ of
$SO(2(n-1))$ when $D_n\to D_{n-1}$.
Since we represent the $D_n$ model as an orbifold of an
$A_{n'}$-model we have to consider the relevant unfolding in the
covering space. Thus, we can use our general framework from section~2.
The generalization from the $A$-series to the $D$-series gives us
in addition the possibility of describing matter in the
antisymmetric representation of $SU(n)$. This is of interest when
studying grand unified models such as $SU(5)$ in which the ${\bf
10}$ plays an important role.

\subsection{Orbifold of A-series gives D-series}

We now use the hyperk\"ahler quotient construction to describe a
$D_n$ singularity.
Because of the non-abelian nature of the corresponding $N=1$ supersymmetric
quiver theory, we will represent the $D_n$ singularity in terms of a
$\mathbb{Z}_2$ orbifold of an $A_n'$ singularity.
In particular, we will focus on the $\mathbb{Z}_2$ invariant untwisted sector.
  From the analysis in appendix B it is sufficient to
consider an $A_{2n-1}$ singularity followed by a $\mathbb{Z}_2$
orbifold. This allows us to represent all of the unfoldings of the $D_n$
HK quotient in
terms of $\mathbb{Z}_2$ invariant deformations of the $A_{2n-1}$ singularity.

To that effect we
start by recalling the set-up for the hyperk\"ahler quotient construction
of an $A_{2n-1}$ singularity as discussed in section~2.
The $A_{2n-1}$ singularity is locally realized as
$\mathbb{C}^2/\mathbb{Z}_{2n}$,
which in turn can be defined as a HK quotient $\mathbb{H}^{2n}//U(1)^{2n-1}$.
The complex scalars $(M_i,\overline{M}_i)\in \Phi_i,\, i=0,\ldots,2n-1$ can
be combined into
gauge invariant baryons and mesons, which from Eq.~(\ref{baryon}) satisfy
\be\label{A2n}
x y = z^{2n}~.
\ee

In order to obtain a $D_n$ singularity we have to further divide by a
$\mathbb{Z}_2$ generator. (For more detail, see appendix~B.) The
action of the
$S$ generator
translates into
\be\label{Z2}
S: \quad (z_1,z_2)\to(z_2,-z_1) \quad:\quad (x,y,z)\to(y,x,-z)~.
\ee
\ie\  $S$ acts non-trivially on the $SU(2n)$ gauge invariant variables.
Following the discussion in appendix~B
new $SO(2n)$ invariant combinations can be defined (see Eq.~(\ref{D-inv-new}))
which satisfies the relation of a $D_n$ singularity~(\ref{Dn-new}).

Next, let us consider a general unfolding of the hyperk\"ahler
quotient describing the $A_{2n-1}$ singularity along the lines of
section~2.  We consider a $\mathbb{Z}_2$ symmetric double unfolding
given by~(\ref{dblunfold}), \ie\ $A_{2n-1}\to A_{2(n-r)}\times
A_{r-1}^2$. In terms of the invariant coordinates~(\ref{barymes})
we have the following relation
\be\label{Dnunfolded}
x\cdot y = z^{2(n-r)} w^r u^r~,
\ee
where $z=z_1 z_2$, $w=z_3 z_4$ and $u=z_5 z_6$.

We can  now use the F-term equations in~(\ref{dblunfold})
to express $w$ and $u$ in terms of
$z$,
\begin{equation}
           w=z+ c_1\,,\quad u=w+c_2=z-c_1\,\quad {\rm with}\quad c_2=-2c_1
\end{equation}
In particular, the relation between the moment maps~(\ref{linconst})
is chosen to reflect the $\mathbb{Z}_2$ symmetric nature of the
configuration\footnote{Note that this choice is different from
that in section~3.}.
Thus, we can rewrite~(\ref{Dnunfolded}) as
\begin{equation}\label{Dnunfoldednew}
           x\cdot y = z^{2(n-r)}(z^2- c_1^2)^r~.
\end{equation}
The $\mathbb{Z}_2$ action in~(\ref{Z2}) now implies that
\begin{equation}
           S:\quad (x,y,z,w,u)\to(y,x,-z,-u,-w)~.
\label{Z2unfold}
\end{equation}
Equivalently, in terms of the $z_i$ we have
\begin{equation}\label{orifold}
    S:\quad (z_1,z_2,z_3,z_4,z_5,z_6) \to
    (z_2,-z_1,z_6,-z_5,z_4,-z_3)~
\end{equation}
which is a symmetry of the D/F term equations~(\ref{hyper1})
if $c_2=-2 c_1$ and $r_2 = -2 r_1$ and can be viewed as the
M theory lift of the orientifold action.
When applied to~(\ref{Dnunfoldednew}) this gives the following
unfolding of the $D_n$ singularity
\be\label{Dn-Dnr}
Y^2=Z X^2 - Z^{-1} Z^{n-r}(Z+t^2)^r\,,\quad t^2=-c_1^2 ~.
\ee

Following Katz and Vafa's analysis for localized matter
hypermultiplets in type IIA \cite{katzvafa} we will now show
that~(\ref{Dn-Dnr}) describes
\be\label{SO2n-SO2nr}
SO(2n)\to SO(2(n-r)\times SU(r)\,,\quad (\bf{2(n-r),r})+(\bf{1,r(r-1)/2})~.
\ee
We first note that for $t=0$ Eq.~(\ref{Dn-Dnr}) describes a $D_n$ singularity.
On the other hand,
when $t\neq 0$ there is a $D_{n-r}$ singularity at $X=Y=Z=0$ while at
$X=Y=Z+t^2=0$ there is an $A_{r-1}$ singularity. The matter, as we
will explain in more detail below, is obtained by decomposing the
adjoint of $SO(2n)$ in terms of representations of $SO(2(n-r))\times
SU(r)$. Note that we keep the chiral multiplets that survive the
breaking of $N=4$ supersymmetry to $N=1$ by the non-trivial fibration
of the ALE space describing the $D_n$ singularity over the base
parametrized by the moment maps.

The fact that the covering space of these spaces can be obtained
by the double unfolding procedure of section 2 allows us to determine
their metric in most cases. The covering space is of the type which
can be related to the twistor space construction, as explained in
section 3, and hence its metric is given by the metric on the
cone over the twistor space of $\mathbb{WCP}^2_{2(n-r-2),r,r}$ for
$n-r > 2$. For $n-r = 2$ the double unfolding Eqs.~(\ref{dblunfold})-
(\ref{hyper1}) degenerates to a simple unfolding \cite{aw1,bobbyed}
and the metric on the covering space is a cone over
$\mathbb{CP}^3/\mathbb{Z}_r$.
Hence, the metric we are interested in is just a $\mathbb{Z}_2$
orbifold~(\ref{orifold}) of a known metric.
However, this requires $n-r \geq 2$ for the following reason:
Although we can represent all $D_n$ singularities as manifolds
in terms of orbifolds of $A_{n^\prime}$ singularities, this is
not true for their metrics. In particular in \cite{chalmers} it was
shown that the M theory lift of
an $\mathcal{O}6^-$ plane with $r$ D6-branes plus mirror
branes is given by a $\mathbb{Z}_2$ orbifold of a multi-center
Taub-NUT metric if at least two D6-branes sit on top of the orientifold
plane. If all D6-branes and the orientifold are together the metric is
an orbifold of a single center Taub-NUT with charge $2 r - 4$ where the
$-4$ comes from the charge of the orientifold plane\footnote{This is
also the reason why the covering space is the cone over the
twistor space of $\mathbb{WCP}^2_{2(n-r-2),r,r}$ and not
$\mathbb{WCP}^2_{2(n-r),r,r}$.}
which is non-singular for $r \geq 2$. On the other
hand a $\mathcal{O}6^-$ plane or $\mathcal{O}6^-$ plane with a single
D6-brane is described
by the Atiyah-Hitchin metric or a $\mathbb{Z}_2$ orbifold of it,
respectively, which are both smooth. In particular this means that
only the cases where at least two D6-branes lie on top of the
$\mathcal{O}6^-$ plane can be related to orbifolds of $G_2$ spaces
from the twistor space construction. These correspond to the
breaking pattern $D_n \to D_{n-r \geq 2} \times A_{r-1}$.

\subsection{Map from M-theory to IIA}
In section~2 (see also appendix~A)
we discussed how compactifying on an $S^1$ maps
the  KK-monopoles in M-theory to D6-branes in
type IIA theory.
In particular, the fixed points under the $U(1)$ action give the
locations of the D6-branes.
The enhanced $SU(n)$ gauge symmetry and charged matter arises from
M2-branes wrapped on 2-cycles in M-theory while in the type IIA
picture open strings stretch between intersecting D6-branes.

This picture gets modified when we consider a
$D_n$ singularity.
Following our earlier discussion we represent the $D_n$ singularity
in terms of a $\mathbb{Z}_2$ orbifold of an $A_n'$ singularity.
In a beautiful paper~\cite{sen} Sen showed that the $\mathbb{Z}_2$ parity
symmetry of the Taub-Nut space, representing the KK-monopoles,
gets mapped to $(-1)^{F_L}
\Omega I$ where $I$ is the parity symmetry in $\mathbb{R}^3$
transverse to the D6-branes. Indeed,
the parity transformation in
M-theory  is exactly the $\mathbb{Z}_2$ given
in~(\ref{Z2}).
We thus have M-theory compactified on an Atiyah-Hitchin
space~\cite{ah} with overlapping
KK monopoles at the origin.
Furthermore, this is in agreement with the $D_n$ singularity described
by~(\ref{Dn}). In type IIA the KK-monopoles get mapped to D6-branes
which are paired by the $\mathbb{Z}_2$ action, while the
Atiyah-Hitchin space corresponds to an $\mathcal{O}6^{-}$ plane.

It is appropriate to make the following remarks here.  For an $A_{2n-5}$
singularity we can formally write
\be\label{TN2n}
x y = z^{-4}z^{2 n} ~.
\ee
This is a semiclassical description of $n$ pairs of D6-branes
located on top of an $\mathcal{O}6^{-}$ plane~\cite{sen}.
Second, to describe the lift of the $\mathcal{O}6^{-}$ plane
to M-theory we need to include
the $S$-generator of the binary dihedral group $\mathbb{D}_{n-2}$,
corresponding to the above $\mathbb{Z}_2$ action. This leads indeed
to the correct description of the
Atiyah-Hitchin space. With the $n$ pairs of D6-branes taken into
account one obtains a $D_n$ singularity,~(\ref{Dn}).

The above picture can be generalized to the unfolding of a $D_n$ singularity.
As in section~2 we start by considering the covering space $A_{2n-1}$
and its deformations. For definiteness let
$A_{2n-1}\to A_{2(n-r)-1}\times A_{r-1}^2$, \ie\ following section~2 we have
an enhanced gauge symmetry $SU(2(n-r))\times SU(r)^2$ with bifundamental
matter $({\bf 2(n-r)},{\bf \overline{r}},{\bf 1})\,+\,({\bf
\overline{2(n-r)}},{\bf 1}, {\bf r})\,+\,({\bf 1},{\bf r},
{\bf \overline{r}})$.
It is now straightforward to read off the gauge symmetry and matter
after the $\mathbb{Z}_2$ transformation~(\ref{Z2unfold}). It is clear
that the $\mathbb{Z}_2$ acts such as $SU(2(n-r))\to SO(2(n-r))$.
Since the two $SU(r)$ are exchanged, see~(\ref{Z2unfold}), this
results in a single $SU(r)$. The matter is read off by
decomposing the adjoint of $SO(2n)$ in terms of $SO(2(n-r))\times
SU(r)$.

\subsection{Exceptional Groups}

We now use the hyperk\"ahler quotient construction to describe an
$E_n$ singularity. In particular, we will focus on $E_6$ and $E_7$.
From our discussion in appendix~C it is sufficient to consider a $D_4$
singularity followed by  a  $\mathbb{Z}_3$ ($\mathbb{Z}_3\times \mathbb{Z}_2$)
orbifold to
construct an $E_6$ ($E_7$) singularity.
Let us start by considering the case of $E_6$.
In order to correctly describe the deformations of the $E_6$
singularity that are inherited as $U$-invariant deformations of the
$D_4$ we find that the correct covering space is the $A_3$
singularity (see Appendix~C). We consider  the single unfolding,
$A_3\to A_1\times A_1$. Following the analysis outlined in section~2,
identifying the appropriate coordinates in terms of the chiral
components $M_i,\overline{M}_i$ of the hypermultiplet $\Phi_i$ we find
\be
x y = w^2 u^2\,,\quad u=w+c~.
\label{A3}
\ee
By a change of variables,
$z=w-e^{i\pi/4} 3^{1/4} t,\, c=-e^{i\pi/4}2\, 3^{1/4} t$, Eq.~(\ref{A3})
becomes $x y = (z^2+ i\sqrt{3} t^2)^2$, where
$z=z_1 z_2$, $w=z_3 z_4$
and $u=z_5 z_6$, $z_i\in \mathbb{C}^6$ (see
section~2).
We show in appendix C that this is a deformation appropriate for
studying the $E_6$
singularity~(\ref{A3def}).
The $U(1)$ action on the $z_i,\, i=3,\ldots, 6$ follows
straightforwardly from section~2 Eq. (\ref{u1action}),
\be
U(1) :  (1/2,-1/2,-1/2,1/2) ~.
\ee
Following the arguments in~\cite{bobbyed}, by rescaling the charges
this corresponds to a $\mathbb{Z}_2$ orbifold of $\mathbb{CP}^3$.
In terms of the projective coordinates $w_i$ of $\mathbb{CP}^3$,
where we identify $w_1=z_3$, $w_2=\bar z_4$, $w_3=z_5$, $w_4=\bar
z_6$, the $\mathbb{Z}_2$ action is given by
\ba\label{Z2orbifold}
\mathbb{Z}_2\,:\, (w_1,w_2,w_3,w_4)\to (-w_1,-w_2,w_3,w_4)~.
\ea
To obtain the unfolding of the $E_6$ singularity
we have to further divide by
$S$ and $U$,
\ba\label{S}
S&:&\quad
(w_1,w_2,w_3,w_4) \to (w_4,-w_3,w_2,-w_1)\\
U&:&\quad (w_1,w_2,w_3,w_4)\to \nonumber \\
&& \quad (\frac{\epsilon}{\sqrt{2}} (w_1-w_4),
\frac{\epsilon^7}{\sqrt{2}} (w_2+w_3),
\frac{\epsilon}{\sqrt{2}} (-w_2+w_3), \frac{\epsilon^7}{\sqrt{2}} (w_1+w_4))~,
\label{U}
\ea
with $\epsilon^8=1$.
The action on the projective coordinates is deduced as in the case of
the unfolding of the $D_n$ singularity as a symmetry of the D/F-term
equations~(\ref{orifold}).

Finally, the matter localized at the singularity is obtained by
decomposing the adjoint of $E_6$, $\bf{78}$, in terms of
representations of $SU(3)^2\times SU(2)\times U(1)$,
\be\label{E6matter}
{\bf (3,\bar 3,1)_2} + {\bf (\bar 3,3,2)_{1}} + {\bf (1,1,2)_{-3}}~.
\ee
Notice the matter charged with respect to all three non-abelian
factors. This is an example in which a perturbative description in
terms of open strings stretching between D6-branes cannot be used
to explain this particular matter content.

For $E_7$ the situation is very similar. The $A_3$ singularity is
once again the covering space, since the deformation given
in~(\ref{A3def}) is invariant under the $V$ transformation.
Hence the unfolding of the hyperk\"ahler quotient proceeds exactly as
for the $E_6$ case above. Thus, the base of the cone is
$\mathbb{CP}^3/(\mathbb{Z}_2 \cdot S\cdot U\cdot V)$, where
$\mathbb{Z}_2$, $S$ and $U$ are given in Eqs.~(\ref{Z2orbifold}),
(\ref{S}) and (\ref{U}), respectively, and
$V$ is given by
\ba\label{V}
(w_1,w_2,w_3,w_4) \to
(\alpha\, w_1,\alpha\, w_2,\alpha\, w_3,\alpha\, w_4)\,,\quad
\alpha^8=1~.
\ea
This action is consistent with the D/F-term equations of a single
unfolding~(\ref{notcnstrnt}).

The matter localized at the singularity is obtained by decomposing
the adjoint of $E_7$, $\bf{133}$ in terms of representations of
$SU(4)\times SU(3)\times SU(2)\times U(1)$~\footnote{The relative $U(1)$
charge assignment is not clear from the hyperk\"ahler quotient
construction. A more detailed study of the anomaly cancellation condition
will hopefully resolve this issue.},
\be\label{E7matter}
{\bf (4, 1,2)_{-3}}+{\bf (6,\bar 3, 1)_{-2}} + {\bf (4,\bar 3,2)_{1}} +
{\bf (1,\bar 3,1)_{4}}~.
\ee
As in the case of $E_6$ there is matter charged with respect to all
three non-abelian factors a reflection of the non-perturbative nature
of the existence of this particular matter content.

There is however no map from M-theory to type IIA in this case. Recall
that for the
D-series it was essential that the parity transformation was
accompanied by $\Omega$. However, $U$ acts like a $\mathbb{Z}_3$
which does not have a well-defined description in type IIA.
Furthermore, the exotic matter in~(\ref{E6matter}) and (\ref{E7matter})
is another reflection of the non-perturbative nature of these models.
A similar situation occurs in
the construction of $E_n$ singularities
in F-theory, in which $\tau$, the axion-dilaton, has to take a
particular value corresponding to a strongly coupled type IIB
theory~\cite{minahan}.

\section{Discussion and Conclusion}

In this paper we studied a large class of singular $G_2$ holonomy
manifolds that give rise to chiral matter. Our construction
is based on a generalization of the simple unfolding of HK quotient
singularities \cite{bobbyed} to multiple unfoldings. We showed
that in general these manifolds, after a reduction along a
suitable circle to type IIA, correspond to configurations
of stacks of D6-branes intersecting over flat $\mathbb{R}^6$.
This gives rise to chiral matter in bifundamental representations
under the gauge group $SU(n_1) \times \ldots \times SU(n_m)$
which can be understood from the open string spectrum in type IIA.
In general the explicit $G_2$ metrics on these spaces are not known
but since the angles between the branes can be chosen to be
supersymmetric this guarantees that a lift of the type IIA
background to a $G_2$ holonomy metric exists. One might wonder where
the angles between the brane enter in our construction. As long
as we do not commit to a metric we only describe the space as
a manifold in Eqs.~(\ref{linconst}) and (\ref{hyper1}) but there
is still a lot of freedom in rescaling the coordinates which
leads to the same manifold. So at this stage, without a metric,
it does not make sense to talk about angles and trying to attribute
them to the numbers appearing in Eq. (\ref{linconst}).
These will eventually be fixed by imposing $G_2$ holonomy but to find
these metrics explicitly will be very hard, except for some of
the examples discussed in section 3 and 4.

In particular in section 3 we were able to map manifolds obtained
via double unfolding to cones over twistor spaces over weighted
projective spaces $\mathbb{WCP}^2_{k_1,k_2,k_3}$ if at least two
of the $k_i$ are equal, say $k_2=k_3$. In this case the metric
and $G_2$ structure can be constructed explicitly \cite{bryant,gibbons}.
In type IIA this describes the intersection of three
sets of D6-branes with $k_1$, $k_2$ and $k_2$ branes in every set.

Furthermore, we extended the possible representations that appear
at the singularities from bifundamental representations of unitary
groups to anti-symmetric representations of unitary groups and
(bi)fundamental representations of orthogonal groups. This was
achieved by introducing $\mathcal{O}6^{-}$ planes into the picture.
In principle this can be done by unfolding $D_n$ singularities,
which is technically harder because the orbifold action is non-abelian.
But as we show in section 4 in many cases we can construct these spaces
as $\mathbb{Z}_2$ orbifolds of double unfolded $A_{2n-5}$ singularities.
For this to work the $\mathcal{O}6^{-}$ plane has to be superimposed with
at least two D6-branes (and their mirrors, if we work in the covering
space). In these cases we can do even more. The $G_2$
metric can be constructed explicitly since it is simply a $\mathbb{Z}_2$
orbifold of spaces arising from double unfoldings which can be related
to the twistor space construction.

We are also able to describe unfoldings of $E_6$ and $E_7$ where
the group is broken to three or more non-abelian factors. (For
these constructions, the metric is also known to have $G_2$ holonomy
since the unfolding
is given by a cone over a non-abelian orbifold of $\mathbb{CP}^3$.)
The corresponding chiral matter fields appear in representations
charged simultaneously under three gauge factors. Obviously, this
has no analogue in weakly coupled string theory and cannot be
explained by open strings stretching between D6-branes.
It would be very interesting to generalize to other breaking
patterns discussed in \cite{katzvafa}, e.g. $E_6 \to SO(10)$ and
$E_7 \to E_6$ which give rise to the ${\bf 16}$ of $SO(10)$
and the ${\bf 27}$ of $E_6$, respectively.

It is definitely interesting and important to get more examples
of singular $G_2$ spaces that produce chiral matter, e.g. recently
a large class of conical  $G_2$ spaces was constructed from a
generalization of the twistor space construction, which automatically
gives the corresponding $G_2$ metric~\cite{calin} (see also \cite{klaus}).
Another challenging task is to find the explicit metrics on the $G_2$
cones constructed via unfolding of HK quotient singularities.
In the case of simple unfolding \cite{bobbyed} the base of the
cone is a weighted projective space $\mathbb{WCP}^3_{n,n,m,m}$
but the corresponding nearly K\"ahler metric is not known.
Similarly, for the general double unfolding with distinct $k_i$ or
multiple unfoldings the metrics are not known.
A possible route to finding them is to exploit the construction
of $G_2$ holonomy metrics from harmonic three-forms $\Phi$.
In this case one starts from an ansatz for the three-form $\Phi$ and
imposing $G_2$ holonomy amounts to solving $d\Phi=d \ast_\Phi \Phi =0$,
which is a highly non-linear differential equation for $\Phi$.
The metric itself is a non-linear function of $\Phi$.
This approach is very efficient and has been employed successfully
in \cite{bryant,brandhuber} to construct complete, non-compact
$G_2$ holonomy spaces.
See also \cite{bggg,cglp2} for a closely related method and
\cite{cglp1} where an effective Lagrangian approach is used.

Our understanding of honest M theory compactification on $G_2$
manifolds is hampered by the lack of a microscopic description
of M theory and the scarcity of examples of compact $G_2$
manifolds. Dualities with string theory imply the existence
of huge classes of compact $G_2$ manifolds, see e.g.
\cite{cvetic,bobbyed}, but up to now only three methods exist,
that can be used to construct them (See \cite{joyce} for a
short review and more references).
In particular most of the $G_2$ singularities that give rise to chiral
matter, which were described in \cite{bobbyed} and this paper, do
not arise in the known examples of compact $G_2$ manifolds.
But since chiral matter is necessary for physically interesting
compactifications more general constructions of compact
$G_2$ manifolds are needed. A related problem is the absence of a simple
topological condition for the existence of a $G_2$ structure on a
given seven-manifold. This should be contrasted with the situation
for Calabi-Yau manifolds where one only has to show the vanishing
of the first Chern class.

\section*{Acknowledgments}
We would like to thank C.~I.~Lazaroiu and E.~Witten for discussions.
P.~B. would like to thank  LBL and in particular
the CIG, Berkeley for  their hospitality.
The work of P.~B.\ was supported in part by  the US
Department of Energy under grant number DE-FG03-84ER40168.
The work of A.~B.\ is supported in part by the DOE under
grant No. DE-FG03-92ER40701. A.~B. would like to thank
Therapy West for hospitality during the final stages of
this research.

\appendix

\section{Reduction to Type IIA}

The singular $G_2$ holonomy manifolds $X$ we constructed
via double (and multiple) unfolding of HK quotient spaces are all
cones over some six-dimensional Einstein manifolds $Y$.
Hence the metric can be written as
\begin{equation}
ds^2 = dr^2 + r^2 d\Omega_Y^2 ~.
\end{equation}
We want to show that $M$-theory compactifications on these
manifolds are the uplift of type IIA backgrounds of
D6-brane and $\mathcal{O}6^{-}$ plane intersecting over flat $\mathbb{R}^6$.
This means that we have to identify a $U(1)$ isometry with the
following properties \cite{aw1}: The
coset $X/U(1)$ is topologically $\mathbb{R}^6$,
however, in general the metric will differ from
flat space, since the dilaton, which is proportional to
the size of the $U(1)$, varies over $\mathbb{R}^6$.
In particular this means that the base of
the cone $Y$ modded by this $U(1)$ action gives $\mathbb{S}^5$ since
$\mathbb{R}^6$ is a cone over $\mathbb{S}^5$.
Furthermore, the $U(1)$ may have
fixed points in $\mathbb{R}^6$ which have the interpretation as the
location of D6-branes/$\mathcal{O}6^{-}$ planes only if they occur in
codimension four of the seven-dimensional $G_2$ manifold.
Since we claim that our manifolds correspond to the intersection of
flat D6-branes/$\mathcal{O}6^{-}$ planes we expect the fixed point sets to be
copies of $\mathbb{R}^3$ inside $\mathbb{R}^6$ and,
because of the conical structure
of $X$, they correspond to copies of two-spheres,
$\mathbb{S}^2=\mathbb{R}^3/U(1)$ inside
$\mathbb{S}^5 = Y/U(1)$.

The hyper-K\"ahler moment map
plays an important role in what follows.
Let us represent $\mathbb{R}^4 = \mathbb{C}^2$
by a complex two-component vector $x = (x_1,x_2)^t$
and introduce the $U(1)$ action $x \to e^{i \theta} x$.
Then the quotient $\mathbb{R}^4 \to \mathbb{R}^4/U(1) =
\mathbb{R}^3$ is given in terms of $U(1)$ invariant moment maps
\begin{equation}\label{moment}
\vec{r} = x^\dagger \vec{\sigma} x \equiv (x, \vec{\sigma} x) ~,
\end{equation}
where $\vec{\sigma}$ are the standard hermitian Pauli matrices.
This map was already introduced in Eq.~(\ref{mommap}) in
component form.

To make contact with our constructions via double (and multiple)
unfoldings Eq.~(\ref{hyper1}) we reorganize the
coordinates of $\mathbb{C}^6$ in three groups of two:
\begin{equation}
x = (z_1,\bar{z}_2) ~,~ y = (z_3,\bar{z}_4)  ~,~ z = (z_5,\bar{z}_6)
\end{equation}
It is then easy to see that the equations in  (\ref{moment}) are just
linear combinations of the moment maps of $x$, $y$ and $z$, and
are invariant under the two $U(1)$ actions (\ref{u1action}).
The $U(1)$ actions on the new coordinates take the form
\be
\label{uonetwo}
(x,y,z) \to (x e^{-i \theta_1/n_1},y e^{i (\theta_1+\theta_2)/n_2},
z e^{-i \theta_2/n_3}) ~.
\ee

In the remaining part of this section we want to show that the $U(1)$ action
\begin{equation}
\label{type2}
(x,y,z) \to (e^{i \theta}x,y,z)
\end{equation}
provides the reduction to type IIA with the desired features listed
above.
First we note that the space indeed has a conical structure
since we can rescale $x,y,z$ by a non-zero real number
without changing the F and D term equations. To construct
the base of the cone we just have to intersect this hypersurface
with the eleven-sphere\footnote{Note, that this describes
an eleven-sphere although the condition is
quartic in the coordinates $z_i$. It is however advantageous for
the rest of the section to use this convention. Since we are only
interested in topology here, we could actually use arbitrary positive
exponents in this equation.}
\begin{equation}\label{11sphere}
(x,x)^2 + (y,y)^2 + (z,z)^2 = 1 ~,
\end{equation}
which avoids the origin in $\mathbb{C}^6$.
Dividing out the three $U(1)$ actions leads to an eight-sphere embedded
in $\mathbb{R}^9$. The eight-sphere can be parametrized by the
nine-vector
\begin{equation}\label{8sphere}
\left( \vec{r}_x, \vec{r}_y, \vec{r}_z \right) =
\left( s \, \vec{e}_x, t \, \vec{e}_y,
\sqrt{1 - s^2 - t^2} \, \vec{e}_z
\right) ~,
\end{equation}
with $s \in [0,1]$, $t \in [0,1]$ and $\vec{e}_x, \vec{e}_y, \vec{e}_z
\in \mathbb{S}^2$. If we define
$\vec{e}_x = (x, \vec{\sigma} x)/(x,x)$,
$\vec{e}_y = (y, \vec{\sigma} y)/(y,y)$,
$\vec{e}_z = (z, \vec{\sigma} z)/(z,z)$,
$s = (x,x)$, $t = (y,y)$, then this gives an isomorphism from
$\mathbb{S}^{11}/U(1)^3$ to $\mathbb{S}^8$.

The D/F-term equations impose a linear relation
between the three terms in (\ref{8sphere}), which for general
D and F-terms takes the form
\begin{equation}\label{linrel}
\vec{r}_x + (R-\mathbb{I}_{3 \times 3}) \cdot \vec{r}_y - R \cdot \vec{r}_z = 0
\end{equation}
where the three by three matrix $R$ was defined in~(\ref{linconst}).
These are the equations
of three hyperplanes intersecting the $\mathbb{S}^8$
defined by $|\vec{r}_x|^2 + |\vec{r}_y|^2 +|\vec{r}_z|^2 = 1$.
The three equations in (\ref{linrel}) are linearly independent,
hence, this gives an $\mathbb{S}^5$ as desired.
The same reasoning can be repeated for multiple unfoldings with
the same result $Y = X/U(1) = \mathbb{S}^5$.

Finally, we want to find the fixpoint set of the $U(1)$ action
(\ref{type2}), which, if it appears at codimension four,
corresponds to the location of D6-branes and/or $\mathcal{O}6^{-}$ planes.
For this we assume that the $\mathcal{O}6^{-}$ plane has at least two
D6-branes on top of it. Otherwise the discussion becomes more
involved since $D_{0,1}$ singularities which correspond to
$\mathcal{O}6^{-}$ planes with zero or one D6-branes on top are
represented by the Atiyah-Hitchin space or a $\mathbb{Z}_2$ orbifold of it,
whereas $D_{n \geq 2}$ singularities are represented by much simpler
$\mathbb{Z}_2$ orbifolds of ALE spaces with $A_{2 n - 5}$ singularities
\cite{chalmers}.
The corresponding $D_n$ ALF metric, if at least two D6-branes coincide
with the $\mathcal{O}6^{-}$ plane, is
just the multicentered Taub-NUT metric with
$2 n - 4$ KK-monopoles arranged in a $\mathbb{Z}_2$
symmetric fashion \cite{chalmers}. The $\mathbb{Z}_2$ is the
M-theory lift of the orientifold action and the multi-Taub-NUT metric
describes the double covering space before modding by the $\mathbb{Z}_2$.

  From the M-theory circle action (\ref{type2}) and the $U(1)^2$ action
given by (\ref{uonetwo}) we see that (\ref{type2}) gives
a fixed point set consisting
of three components: $x = 0$, $y=0$ and $z=0$.
On any of these components Eq.~(\ref{11sphere}) reduces to the equation for
an $\mathbb{S}^7$. Out of the three $U(1)$ actions one particular
linear combination acts trivially,
so that dividing out the $U(1)$ symmetries gives an $\mathbb{S}^5$.
Finally, we have to impose the three linear relations (\ref{linrel})
which yield an $\mathbb{S}^2$ as expected. The fixed point set is the
union of three $\mathbb{S}^2$'s, which do not intersect
except at the tip of the cone where the
whole base shrinks to a point. Hence, in type IIA this corresponds to
three sets of intersecting D6-branes. To find the multiplicities
of the D6-branes we have to identify the type of singularity at
the fixed points. Inspecting the $U(1)$ actions we find
an $A_{n_1-1}$ singularity at $x=0$, an $A_{n_2-1}$ singularity at $y=0$
and an $A_{n_3-1}$ singularity at $z=0$, which means that we have
three stacks of D6-branes with multiplicities $n_i ~,~i=1,2,3$.

It is important to note that the codimension four singularities all
coincide with the $U(1)$ fixed point set. Only for this reason do we have
a clean interpretation in terms of D6-brane configurations.
In the twistor construction (see also \cite{calin}) this is not the
case in general and the interpretation is more complicated which
makes it hard to identify the correct spectrum of chiral fields at
the singularity \cite{bobbyed,calin}.
Also, most of the arguments go through for multiple unfoldings
under relatively mild assumptions for the generalization of
Eqs.~(\ref{hyper1}) and (\ref{linrel}).
Concretely, this means for an $m$-fold unfolding that there should
be $3 m-3$ linearly independent equations.
In this case the corresponding $G_2$ singularity corresponds
to the intersection of $m+1$ stacks of D6-branes.

\section{Construction of $D_n$ singularities}

$D_n$ singularities can be constructed by dividing the complex
two-plane $\mathbb{C}^2$ by a finite non-abelian group,
the binary dihedral group $\mathbb{D}_{n-2}$ of rank $4 (n-2)$,
generated by
$T =
\left( \begin{array}{cc}
\xi & 0 \\
0 & \xi^{-1}
\end{array}
\right)$
with $\xi^{n-2} = 1$, and
$S =
\left( \begin{array}{cc}
0 & -1 \\
1 & 0
\end{array}
\right)$ acting on $(z_1 , z_2)\in \mathbb{C}^2$~\cite{slodowy}.
The well-known form of the singularities as equations in $\mathbb{C}^3$
can be found by introducing invariant monomials in terms of the
coordinates $z_1 , z_2$.

We will present this in a two step process where we first divide
by the $\mathbb{Z}_{2(n-2)}$ subgroup generated by $T$,
followed by the $\mathbb{Z}_2$ generated by $S$.
A $T$-invariant set of variables is given by
\begin{equation}
x = z_1^{2(n-2)} \, , \, y = z_2^{2(n-2)} \, , \, z = z_1 z_2
\end{equation}
yielding the $A_{2n-5}$ singularity
\begin{equation}\label{A2n-5}
x y = z^{2(n-2)} ~.
\end{equation}

The fully invariant set of variables is
\begin{equation}\label{D-inv}
\begin{array}{rcl}
Y & = & \frac{1}{2} ( x + y ) = \frac{1}{2}(z_1^{2n-4} + z_2^{2n-4})\\
X & = & \frac{1}{2} z ( x - y ) = \frac{1}{2} z_1 z_2
(z_1^{2n-4} - z_2^{2n-4})\\
Z & = & z^2=z_1^2 z_2^2
\end{array}
\end{equation}
which yields the equation for the $D_n$ singularity
\begin{equation}\label{Dn}
X^2 = Z Y^2 - ZZ^{n-2}  ~.
\end{equation}
Note that in deriving~(\ref{Dn}) we have in the last term used the
$A_{2n-5}$ relation~(\ref{A2n-5}) above.

It is clear from~(\ref{Dn}) that there are only $n-2$ $\mathbb{Z}_2$
($S$)-invariant deformations of~(\ref{A2n-5}), while a $D_n$ singularity
has $n$ deformations.  However, this problem can be circumvented.
We start by considering $\mathbb{Z}_{2n}$ generated by $T$, for which
\begin{equation}
x = z_1^{2n} \, , \, y = z_2^{2n} \, , \, z = z_1 z_2
\end{equation}
yield the $A_{2n-1}$ singularity
\begin{equation}\label{A2n-1}
x y = z^{2n} ~.
\end{equation}
Then, introduce a new set of invariant variables (under the $T$ and
$S$ generators)\footnote{This map is not valid at $x,y\neq 0,\, z=0$. As
we argue below, there are certain limits when $x,y,z$ all vanish in
which~(\ref{D-inv-new}) is still well-defined.}
\begin{equation}\label{D-inv-new}
\begin{array}{rcl}
Y & = & \frac{1}{2} z^{-2}( x + y ) = \frac{1}{2}(z_1
z_2)^{-2}(z_1^{2n} + z_2^{2n})\\
X & = & \frac{1}{2} z^{-1} ( x - y ) = \frac{1}{2} (z_1 z_2)^{-1}
(z_1^{2n} - z_2^{2n})\\
Z & = & z^2=z_1^2 z_2^2
\end{array}
\end{equation}
which also yields the equation for a $D_n$ singularity
\begin{equation}\label{Dn-new}
X^2 = Z Y^2 - Z^{-1}Z^{n}  ~.
\end{equation}
However, the difference is that all of the deformations
of~(\ref{Dn-new}) can be accounted for when deforming~(\ref{A2n-1})
with $S$-invariant deformations,
\ba\label{atod}
x y &=& \prod_{i=1}^n (z^2 - z_i^2)\,\quad \to \\ 
X^2 &=& Z Y^2 - Z^{-1} \Big(\prod_{i=1}^n (Z + Z_i)-\prod_{i=1}^n
Z_i\Big) +2 Y \prod_{i=1}^n Z_i ~,\label{AtoD}
\ea
and the map~(\ref{D-inv-new}) becomes
\begin{equation}\label{D-inv-new-def}
\begin{array}{rcl}
Y & = & \frac{1}{2} z^{-2}( x + y + 2 \prod_{i=1}^n z_i) \\
X & = & \frac{1}{2} z^{-1} ( x - y ) \\
Z & = & z^2~.
\end{array}
\end{equation}

This argument agrees with the general idea that when the D6-branes
are located away from the $\mathcal{O}6^{-}$ plane
they do not feel its effect, and
the theory is perfectly described by $n$ pairs of
D6-branes~\cite{sen}.
However, we have to be careful when considering the  limit $Z_i\to
0$, \ie\ is the map between the covering space of deformations of the
$A_{2n-1}$ singularity
and the space of deformations of the $D_n$ singularity well-defined
in this limit?
For our purpose it is enough to analyze limits of deformations corresponding to
\be\label{Dnbreaking}
D_n\to D_{n-r}\times A_{r-1},
\ee
as they are the ones relevant for studying the localized matter.
Hence, let $z_i^2=-t^2\,,\, i=1,\ldots,r$. In the $A_{2n-1}$ covering
space we get from~(\ref{atod})
\be\label{Anr}
x y = z^{2(n-r)} (z^2+t^2)^r
\ee
while the $D_n$ deformation from~(\ref{AtoD}) is similarly given by
\be\label{Dnr}
X^2 = Z Y^2 - Z^{-1}\Big(Z^{n-r}(Z+t^2)^r - \delta_{n,r} t^{2n}\Big)+
2\delta_{n,r} t^{n} y~.
\ee

Clearly, when $t\neq 0$ we have a $D_{n-r}$ singularity at $X=Y=Z=0$
while at $X=Y=Z-t^2=0$ we have an $A_{r-1}$ singularity. In the
covering space this corresponds to an $A_{2(n-r)-1}$ singularity at
$x=y=z=0$ while at $x=y=z\pm i t=0$ we have two $A_{r-1}$
singularities respectively.
Thus, for $t\neq 0$ we have a well-defined map from the
$\mathbb{Z}_2$ invariant deformations of $A_{2n-1}$ to the
deformations of $D_n$ given by~(\ref{D-inv-new-def}).

Let us now consider the limit $t\to 0$. From~(\ref{Dnr}) we find
a $D_n$ singularity at $X=Y=Z=0$. In the covering space, both $x$ and
$y$ vanish at the location of the singularities when $t\neq 0$. Thus,
in spite of the apparent singular map due to the
negative powers of $z$ in~(\ref{D-inv-new-def}), the
limit $t\to 0$ is well-defined also in the covering space, and
\be\label{Ant0}
A_{2(n-r)-1}\times (A_{r-1})^2\to A_{2n-1}~.
\ee
We can therefor extend the map between the deformation spaces to also
include the origin in the limit $t\to 0$.\footnote{There are other
limits in which the origin can be included but this suffices for our
discussion.}

Finally, note that for $r=n$ the $t\neq 0$ deformation corresponds to
$D_0\times A_{n-1}$, where $D_0$, the Atiyah-Hitchin space, is
located at $X=Y=Z=0$ and the $A_{n-1}$ singularity is at
$X=Y-t^{2(n-2)}=Z+t^2=0$.
Just as for $r<n$ the limit $t\to 0$ is well-defined; $D_0\times
A_{n-1}\to D_n$ and in the covering space $A_{n-1}^2\to A_{2n-1}$.

The $D_0$ singularity is a smooth configuration obtained from the
$D_1$ singularity
by a $\mathbb{Z}_2$ orbifold $(X,Y,Z) \to (-X,-Y,Z)$, where
the $X,Y,Z$ are given in terms of the $A_1$ covering space
coordinates $x,y,z$ in~(\ref{D-inv-new}) for $n=1$.
Introducing new variables $\tilde{Y} = Y^2 , \tilde{X} = X Y$
we find
\be\label{D0}
\tilde{X}^2 = Z \tilde{Y}^2 - \tilde{Y} ~.
\ee
Indeed,~(\ref{D0}) is non-singular.

\section{Construction of $E_n$ singularities}
The $E_n$ singularities can be constructed by extending the binary
dihedral group action, $\mathbb{D}_2$, on the complex two-plane,
$\mathbb{C}^2$~\cite{slodowy}.
Consider the following actions on $(z_1,z_2)\in\mathbb{C}^2$
\ba\label{UV}
U = {1\over\sqrt{2}}
\left( \begin{array}{cc}
\epsilon^7 &  \epsilon^7\\
\epsilon^5 & \epsilon
\end{array}
\right)
     ~~ \mathrm{and} ~~
V =
\left( \begin{array}{cc}
\epsilon & 0\\
0 & \epsilon^7
\end{array}
\right)~,
\ea
where $\epsilon^8=1$. The
binary tetrahedral group, $\mathbb{T}$ of order 24
corresponding to $E_6$ is generated by adding $U$ to $\mathbb{D}_2$.
The $E_7$ singularity is obtained by adding
$V$
to $\mathbb{T}$ which gives the binary octahedral group of order 48.

Let us first consider the $E_6$ singularity. In terms of the $D_4$
invariant variables $X,Y,Z$ we have the following set of variables
invariant under $T$, $S$ and $U$,
\begin{equation}\label{E6var}
\begin{array}{rcl}
\tilde X & = & Y(Y^2-9 Z^2)\\
\tilde Y & = & 3 Z^2 + Y^2\\
\tilde Z & = & (-108)^{1/4}X
\end{array}
\end{equation}
which gives the equation for an $E_6$ singularity
\be\label{E6}
\tilde X^2 = \tilde Y^3 +{\tilde Z^4\over 4}~.
\ee

We are interested in the deformation space associated to the $E_6$
singularity. In analogy with the $D_n$ singularities we want to
describe this parameter space in terms of the covering space of
deformations of the $A_3$ singularity. However, it is not possible to
find an auxiliary representation in which all the deformations of
$E_6$ can be represented. Since the
$U$-invariant deformation also has to be invariant under the $S$
transformations of the $A_3$ singularity we are left with only one
invariant deformation,
\be
x y = (z^2 + i\sqrt{3} t^2)^2
\label{A3def}
\ee
which is mapped to the following deformation of the $E_6$ singularity
\be
\tilde X^2 = \tilde Y^3 + \frac{\tilde  Z^4}{4} - 3 t^2 \tilde Y
\tilde Z^2 + 9 t^4 \tilde Y^2 - 4 t^6 \tilde Z^2 + 24 t^8 \tilde Y +
16 t^{12}~.\label{E6-deformation}
\ee

Clearly, when $t=0$~(\ref{E6-deformation}) describes an $E_6$
singularity. When $t\neq 0$ it is straightforward to show that there
are three different singularities located at $(\tilde X,\tilde
Y,\tilde Z)=(0,0,\pm \sqrt{2} t^3)$ and $(0,-4 t^4,0)$ corresponding
to $A_2$ and $A_1$ type singularities respectively. 
Thus, the deformations in~(\ref{E6-deformation}) correspond to the following
breaking of $E_6$
\be
E_6 \to A_2^2 \times A_1~.\label{E6-breaking}
\ee

Let us now turn to the $E_7$ singularity. We can express the $D_4$
invariant variables in combinations that are invariant under $T$,
$S$, $U$ and $V$,
\begin{equation}\label{E7var}
\begin{array}{rcl}
\hat X & = & X Y(Y^2-9 Z^2)\\
\hat Y & = & 48^{1/3}(3 Z^2 + Y^2)\\
\hat Z & = & 3 X^2
\end{array}
\end{equation}
which gives the equation for an $E_7$ singularity
\be\label{E7}
\hat X^2 = \hat Z \hat Y^3 +16 \hat Z^3~.
\ee

As for $E_6$ we are interested in the deformation space associated to
the $E_7$ singularity. The $U$ and $S$-invariant deformation of the
$A_3$ singularity is also
invariant under the $V$-transformation, and gets mapped to
the following $E_7$ deformation
\be
\hat X^2 = \frac{\hat Y^3\hat Z}{16} -Z^3 + 12 t^2 \hat Y \hat Z^2
-36 t^4 \hat Y^2 \hat Z - 4 t^6 \hat Z^2 + 24 t^8 \hat Y \hat Z - 4
t^{12} \hat Z~.\label{E7-deformation}
\ee
Clearly, when $t=0$~(\ref{E7-deformation}) describes an $E_7$
singularity. When $t\neq 0$ it is straightforward to show that there
are three different singularities located at $(\tilde X,\tilde
Y,\tilde Z)=(0,t^4,0)$, $(0, t^4/4,0)$ and $(0,0,-2 t^6)$
corresponding to $D_3 \equiv A_3$, $A_2$ and $A_1$ type singularities
respectively. Thus, the deformations in~(\ref{E7-deformation})
correspond to the following breaking of $E_7$
\be
E_7 \to (D_3 \equiv A_3) \times A_2 \times A_1~.\label{E7-breaking}
\ee

\end{document}